\pdfoutput=1	%
\documentclass[9pt,twocolumn,twoside]{pnas-new}
\setboolean{displaywatermark}{false}

\templatetype{pnasresearcharticle} %

\title{Probabilistic Value-Deviation-Bounded Integer Codes for Approximate Communication}

\author[1]{Phillip Stanley-Marbell}
\author[2]{Paul Hurley} 

\affil[1]{Department of Engineering, University of Cambridge, Cambridge CB3 0FA, UK}
\affil[2]{IBM Research\,---\,Z\"{u}rich, S\"{a}umerstrasse 4, CH-8803 R\"{u}schlikon, Switzerland}

\leadauthor{Stanley-Marbell} 

\authorcontributions{PSM formulated the problem, derived the bounds, and developed the encoding. PH developed the improved algorithm for generating the code tables. Both authors contributed to writing.}
\correspondingauthor{\textsuperscript{1}To whom correspondence should be addressed. E-mail: {phillip.stanley-marbell@eng.cam.ac.uk}}

\vspace{-0.1in}

\usepackage{etoolbox}	%
\newtoggle{vdbeTightFormatting}
\togglefalse{vdbeTightFormatting}
\newtoggle{pnasTemplate}
\toggletrue{pnasTemplate}
\usepackage{subfig}
\usepackage{xspace}
\usepackage{bm}
\usepackage{graphicx}
\usepackage{grffile}	%
\usepackage{textcomp}
\usepackage{color}
\usepackage{colortbl}
\usepackage{booktabs}
\usepackage{subeqnarray}
\usepackage[ruled,vlined]{algorithm2e}
\usepackage{amsfonts}
\usepackage{amssymb}
\usepackage{amsmath}
\usepackage[thmmarks,amsmath,hyperref]{ntheorem}
\usepackage{xfrac}		%
\usepackage{moreverb}	%
\usepackage{framed}		%

\usepackage{pifont}				%
\usepackage[scaled]{beramono}
\usepackage[T1]{fontenc}

\definecolor{shadecolor}{rgb}{0,0,0}%
\definecolor{greyout}{rgb}{0.65,0.7,0.7}%
\definecolor{tbhead}{rgb}{0.97,0.97,0.97}%
\definecolor{a}{rgb}{0.9,0.95,0.95}%
\definecolor{b}{rgb}{0.99,0.99,0.99}%
\definecolor{white}{rgb}{1, 1, 1}%

\def \logicone {{\small\tt 1}\xspace}
\def \logiczero {{\small\tt 0}\xspace}

\newcommand\Warp					{Warp\xspace}

\newcommand{\Hair}{\ifmmode\mskip1mu\else\kern0.08em\fi}

\newcommand\encodingRadix			{r}
\newcommand\integerDistance			{m}
\newcommand\errorFreeValue			{s}
\newcommand\errorContainingValueSet	{S}
\newcommand\polynomial					{\pi}

\newcommand\wordLength				{L}
\newcommand\yKstarLm				{y_{k^*}(\wordLength, m)}
\newcommand\zKstarLm				{z_{k^*}(\wordLength, m)}
\newcommand\zLkm					{z(\wordLength, k, m)}

\newcommand{\Z}{\mathbb{Z}}
\newcommand{\abs}[1]{\left|#1\right|}

\theoremstyle{plain}
\theoremsymbol{\ensuremath{\diamondsuit}}
\theoremseparator{.}
\theoremprework{\bigskip\hrule}
\theorempostwork{\hrule\bigskip}
\newtheorem{Definition}{Definition}

\begin{abstract}
When computing systems can tolerate the effects of errors or erasures
in their communicated data values, they can trade this tolerance
for improved resource efficiency.  One method for enabling this
tradeoff in the I/O subsystems of computing systems, is to use
channel codes that reduce the power needed to send bits on a channel
in exchange for bounded errors and erasures on numeric program
values---\textit{value-deviation-bounded (VDB)} codes. Unlike rate
distortion codes, which guarantee a bound on the expected value of
channel distortion, the probabilistic VDB codes we present guarantee
any desired tail distribution on integer distances of words transmitted
over a channel.

\qquad \iftoggle{pnasTemplate}{We extend prior work to present tighter upper bounds on the
efficiency for VDB codes.}{}
We present a new probabilistic VDB encoder that lowers power
dissipation in exchange for bounded channel integer distortions.
The code we present takes the peculiar approach of changing the
channel bit error rate across the ordinal bit positions in a word
to reduce power dissipation. We implement the code table generator
in a software tool built on the dReal SMT solver and we validate
the generated codes using Monte Carlo simulation. We present one
realization of hardware to implement the technique, requiring
2\,mm$^2$ of circuit board area and dissipating less than 0.5\,$\mu$W.

\iftoggle{pnasTemplate}{\qquad }
\end{abstract}

\dates{This manuscript was compiled on \today}

\begin{document}

\verticaladjustment{-5pt}

\maketitle
\thispagestyle{firststyle}
\ifthenelse{\boolean{shortarticle}}{\ifthenelse{\boolean{singlecolumn}}{\abscontentformatted}{\abscontent}}{}

\vspace{-0.1in}
\dropcap{M}{ost} computing systems are designed to prevent errors in computation,
memory, and communication.  Guarding against errors however requires
energy, temporal redundancy, or spatial redundancy and therefore
consumes resources.  But not all systems need to be free of errors:
In some systems, either by explicit design or by the nature of the
problems they solve, system output quality degrades gracefully in
the presence of errors.

\iftoggle{pnasTemplate}
{
Several important future applications of computing systems, ranging
from wearable health-monitoring systems to neuromorphic computing
architectures~\cite{merolla2014million} often dissipate a significant
fraction of their energy communicating data (e.g., from sensors)
in which the effects of errors are best quantified in terms of their
integer distance, rather than using a Hamming distance.  At the same time,
the computations that consume this data can often incur errors with
a wide range of integer distances with limited system-level
consequences.
}
{}

Because electrical communication interfaces do not benefit as much
as processors and other digital logic from semiconductor process
technology scaling, the fraction of system energy dissipated by
sensor data access and communication is currently
large~\cite{hatamkhani2006power, mahony2009future, Malladi:2012}
and will only grow in future embedded sensor-driven \iftoggle{pnasTemplate}{and neuromorphic}{}
systems. It is therefore an important challenge to find ways to
reduce the power dissipated in sensor data acquisition and transfer.
One way to do this is by exploiting knowledge of the properties and
uses of the transported data, to reduce communication power dissipation
in exchange for, e.g., bounded inaccuracy of the communicated
data~\cite{Stanley-Marbell:2016:RSI:2897937.2898079, stanley2015efficiency,
kim2017axserbus, stanley2016encoder, huang2015acoco}.

Reducing I/O power in exchange for bounded inaccuracy requires a physical
channel property that provides such a tradeoff. One example of such
a tradeoff is between energy dissipated in the pull-up resistors
on a serial communication interface such as I2C and the link error
or erasure rate. The choice of the pull-up resistor value influences
signal integrity at a given I/O speed and at the same time affects
communication power dissipation.

\begin{figure}[t]
\centering
\includegraphics[trim=0cm 0cm 0cm 0cm, clip=true, angle=0, width=0.425\textwidth]{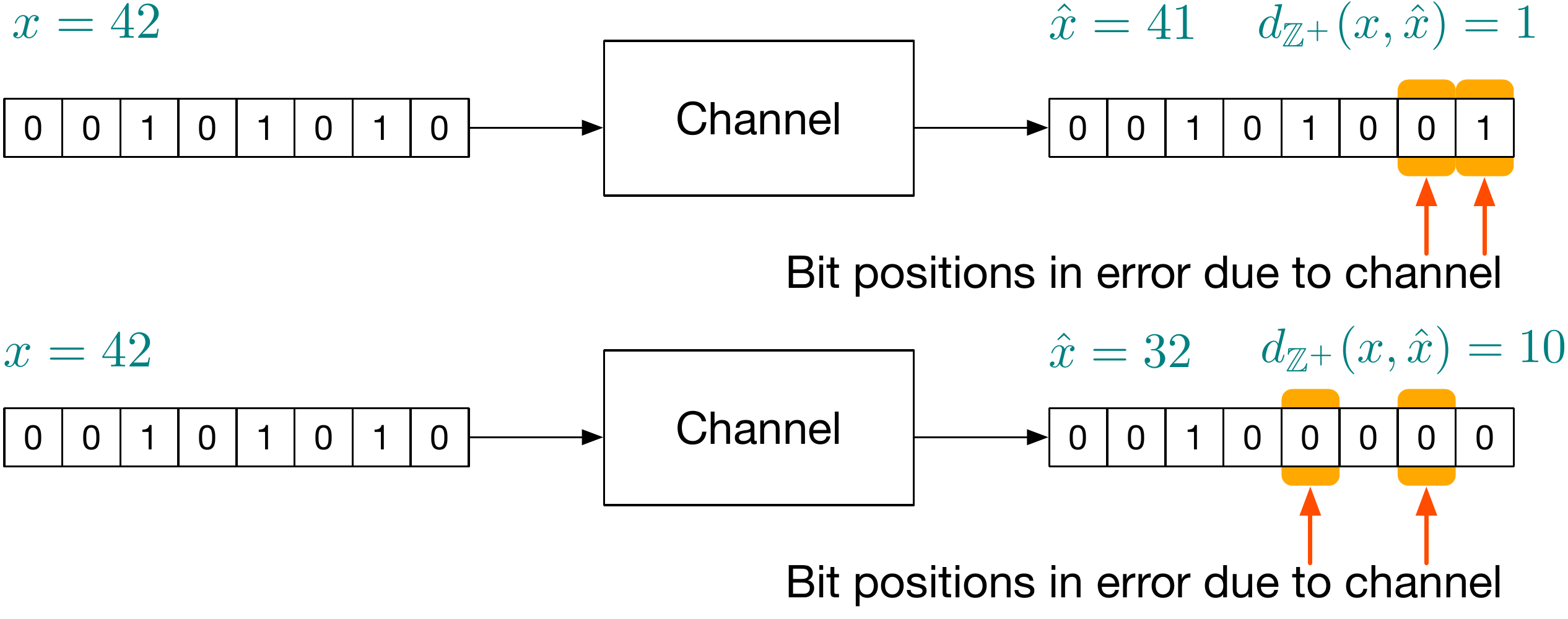}
\vspace{-0.1in}
\caption{Channel model and distance functions: The same number
of bit errors (same Hamming distance) can lead to vastly 
different value deviations (integer distances).}
\label{fig:channel-model}
\end{figure}

\subsection{Contributions}
We exploit value representation and hardware properties to design
codes that trade lower power dissipation for a given bound on the
tail distribution of channel integer distortion.

\iftoggle{pnasTemplate}{
\textbf{Tighter bounds on properties of VDB codes
(Section~\ref{section:bounds})}: We extend prior
work~\cite{StanleyMarbell:itw09} to present tighter upper bounds
on the efficiency of VDB codes.
}{}

\textbf{A new encoding technique for approximate communication: Probabilistic VDB channel encoding
(Section~\ref{section:encoding}):} In a system with integer random
variable input $X$ and output $\hat{X}$, let the random variable
$M=d_{\Z+}(X,\hat{X})$ denote the absolute value of the difference
between $X$ and $\hat{X}$. Figure~\ref{fig:channel-model} shows an
example.  The random variable $M$ takes on instance values $m$, and
has tail distribution $\Pr(M > m)$. In this article, we denote integer
distance errors tolerated by a given end-to-end application
using an upper bound $\widehat{\mathrm{F}}_{\!M}(m)$ on the tail
distribution of $m$ seen at the output of the system (e.g., the
receiver in Figure~\ref{fig:channel-model}). We present a method
to derive an application-specific encoding given any instance of
the function $\widehat{\mathrm{F}}_{\!M}(m)$.  The method we present
generates a custom encoding such that $\Pr(M > m) \le
\widehat{\mathrm{F}}_{\!M}(m)$ for any $\widehat{\mathrm{F}}_{\!M}(m)$
that is a monotone nondecreasing function of $m$ with range $[0,
1]$.  Because it bounds the entire tail distribution
$\widehat{\mathrm{F}}_{\!M}(m)$ of integer distortion, the technique
we present is a generalization of traditional rate distortion codes:
rate distortion codes typically only bound the expected value,
$E[M]$, of the channel distortion. We present our new encoding
approach for identically-distributed channel errors
(Section~\iftoggle{pnasTemplate}{3\,}{}\ref{section:iidBernoulli})\iftoggle{pnasTemplate}{,}{ and} non-identically-distributed
channel errors (Section~\iftoggle{pnasTemplate}{3\,}{}\ref{section:nonIidErrors})\iftoggle{pnasTemplate}{, and we
present an analytic formulation for the effect of the distribution
of transmitted values on masking (Section~3\,\ref{section:maskedErrors}).}{.}

\textbf{Algorithms for probabilistic VDB encoding
(Section~\ref{section:implementation}):} We first present an
easy-to-implement method for generating probabilistic VDB code
tables for a given bound on integer distortion.  Because it only
needs to be run once, offline, when a system is being designed, the
first method we present is practical for the sensor sample resolutions
common in embedded sensor systems (e.g., 8--16 bits).  We then use
techniques borrowed from finite-field arithmetic to create an
efficient algorithm for generating the code tables.

\textbf{A custom hardware implementation (Section~\ref{section:hardware}):}
We demonstrate the feasibility of probabilistic VDB encoding by describing
our implementation of the required hardware support in a custom printed
circuit board platform.

\iftoggle{pnasTemplate}{
Traditional \textit{source coding} changes the words being transmitted
on a channel to adapt to properties of the inputs (e.g., for
compression). Traditional \textit{channel coding} changes words
being transmitted to adapt to properties of the channel (e.g., for
forward error correction). In contrast, the encoding technique we
present in this article (probabilistic VDB coding) takes the peculiar
approach of changing the properties of the communication channel
for different ordinal positions in a word to be transmitted.
Probabilistic VDB coding does this with the objective of reducing
power dissipation by inducing channel errors that lead to values
at the channel's receiver which are at small integer distances from
those sent by the channel's transmitter.
}{}

\subsection{Related Research}
\label{section:relatedwork}
Value-deviation-bounded (VDB) integer codes~\cite{StanleyMarbell:itw09}
were proposed as a means of trading energy efficiency for correctness
in the context of program variable encodings when program variables
can tolerate some distribution of integer value distortions. For
serial communication interfaces such as I2C and SPI which are common
in energy-constrained embedded systems, serial VDB codes (VDBS
encoding)~\cite{stanley2015efficiency,
Stanley-Marbell:2016:RSI:2897937.2898079} provide a concrete encoding
method.  VDBS encoding is deterministic and has the effect of
re-quantizing the number representation of values. This deterministic
distortion may be undesirable for certain applications: When applied
to images with encoder settings to maximally reduce power dissipation,
VDBS encoding leads to regular quantization banding in images.  For
these and similar applications, encoding methods that trade
energy-efficiency for accuracy and whose induced distortion is
stochastic are therefore desirable.

Rate distortion theory~\cite{Shannon:59,Shannon:inftheorybook}
investigates the tradeoff of encoding efficiency ({\it rate}) for
deviation of encoded values ({\it distortion}). A {\it distortion
function} or {\it distortion measure} specifies this distortion as
a function of the original data and its encoded form.  The value
deviations we consider in this work are integer distance distortions.
Unlike the convention in rate distortion theory, the encoding we
present guarantees bounds on the entire distribution of distortions,
rather than only guaranteeing bounds on the expected value of the
distortion.

\iftoggle{pnasTemplate}{
\begin{table}[t]
\caption{Terminology and notation.}
\iftoggle{vdbeTightFormatting}{\vspace{-0.1in}}{}
\begin{tabular}{lll}
\toprule
{\bf Notation}			&	{\bf Definition}						&	{\bf Example}\\
\hline
\rowcolor{a}$\wordLength$&	Word length	(in bits)					&	8\\
			$x$			&	Error-free value						&	42 ({\scriptsize\tt 0010{\color{greyout}1}0{\color{greyout}1}0})\\
\rowcolor{a}$\hat{x}$	&	Value after channel error				&	32 ({\scriptsize\tt 0010{\color{greyout}0}0{\color{greyout}0}0})\\
			$m$			&	Value deviation, $\vert \hat{x}-x\vert$	&	10\\

\rowcolor{a}$M$			&	Random variable denoting				&	\\
\rowcolor{a}			&	value deviations						&	\\
			$\mbox{F}_{\!M(m)}$&	Tail distribution, $\Pr(M > m)$	&	$\begin{cases} 0 & m < 4\\ \sfrac{1}{100} & m=4\\ 1 & m \ge 10\end{cases}$\\
\rowcolor{a}$k$			&	Number of bits of $x$ perturbed in $\hat{x}$&	2 (grayed-out above)\\
\bottomrule
\end{tabular}
\label{table:definitions}
\vspace{-0.05in}
\end{table}
}{}

\section{Definitions}
\label{section:definitions}
Consider an $\wordLength$-bit binary sequence $x \in \Z_2^\wordLength$ transmitted on
a channel. Let the channel output be $\hat{x} \in \Z_2^\wordLength$ and let
$d(x, \hat{x})$ be the Hamming distance between channel input $x$ and
output $\hat{x}$.

In this work we consider systems in which the channel errors are
defined by considering $x$ as an unsigned integer.  Let $x_i$ denote
the $i$'th bit of $x$ and let  $u: \Z_2^\wordLength \rightarrow \Z$
denote the conversion to an integer, with
\begin{align*}
u(x)=\sum_{i=0}^{\wordLength-1} x_i 2^i .
\end{align*}
Then, for unsigned integer value representations of the channel
input and output, we define the distance $d_{\Z^+}(x, \hat{x})$
between channel input and output as
\begin{align*}
d_{\Z^+}(x, \hat{x})=\abs{u(x)- u(\hat{x})} .
\end{align*}

\iftoggle{pnasTemplate}{Table~\ref{table:definitions} summarizes the notation we use in the
rest of this work.}{}  The random variable $M$ denoting values of
channel distortion, with instance values $m$, takes on values
determined by the distance function $d_{\Z^+}(x, \hat{x})$. In the
following, we will use the variable $m$ when we wish to refer to
specific instances of channel integer distortion, and we will use
$d_{\Z^+}(x, \hat{x})$ when we want to emphasize the distortion
function.

Following prior work~\cite{StanleyMarbell:itw09}, the absolute value of the
distance of words transmitted on a channel, when $k$ bit-level
errors for $\wordLength$-bit words can occur, is in the range
\begin{align}
m_{\mathrm{min}} \quad \le \quad  d_{\Z^+}(x, \hat{x}) \quad \le \quad 2^{\wordLength - k}\left( 2^k - 1 \right) \label{eqn:maxm} .
\end{align}
The lower bound on integer distance, $m_{\mathrm{min}}$ is $1$ for a binary
symmetric channel, and is $2^k - 1$ for the binary asymmetric
channel. The upper bound on $m$ in both symmetric and asymmetric
channels occurs when all $k$ channel errors are of the same polarity
(e.g., \logicone~$\rightarrow$~\logiczero) and occur in the most
significant $k$ bit positions. The lower bound on $m$ for asymmetric
channels errors occur when all $k$ channel errors are in the least
significant $k$ bits. For the binary symmetric channel, the lower
bound on $m$ occurs when all channel errors occur in consecutive
bit positions, with the least significant $k - 1$ of the channel
errors having negative polarity (i.e., \logicone~$\rightarrow$~\logiczero),
and the most significant channel error bit position having positive
polarity (i.e., \logiczero~$\rightarrow$~\logicone). Because the low-speed I/O
links used in many sensor-driven systems do not employ any of the
sophisticated channel encodings used for high-speed I/O links, we
restrict our analysis in this work to binary channels.

We use a combinatorial formulation to specify the syndrome of the
codes we develop in Section~\ref{section:encoding}. Our combinatorial
analysis is centered on the number of unique placements of $k$-bit 
channel errors into an $\wordLength$-bit word for a given
value of the integer distortion.

\begin{figure*}
\centering
\iftoggle{pnasTemplate}{\begin{minipage}[b]{0.725\textwidth}}{}
\iftoggle{pnasTemplate}{\begin{framed}}{}
\iftoggle{pnasTemplate}{\vspace{-0.2in}}{}
\begin{align*}
\begin{split}
& \mbox{When up to $k$ upsets can occur in a word, there are only $\yKstarLm$ unique channel}\\
& \mbox{error vectors which can lead to a value deviation of $m$.}\\
m=1 : \quad &\{\stackrel{\mbox{\tiny\tt 000}}{v=0},\stackrel{\mbox{\tiny\tt 00}\stackrel{\bullet}{\mbox{\tiny\tt 1}}}{w=1}\},
		\{\stackrel{\mbox{\tiny\tt 001}}{v=1},\stackrel{\stackrel{\ \bullet\bullet}{\mbox{\tiny\tt 010}}}{w=2}\},
		\{\stackrel{\mbox{\tiny\tt 010}}{v=2},\stackrel{\stackrel{\ \ \bullet}{\mbox{\tiny\tt 011}}}{w=3}\},	
{\color{greyout}\{\stackrel{\mbox{\tiny\tt 011}}{v=3},\stackrel{\stackrel{\bullet\bullet\bullet}{\mbox{\tiny\tt 100}}}{w=4}\},}	
		\{\stackrel{\mbox{\tiny\tt 100}}{v=4},\stackrel{\stackrel{\ \ \bullet}{\mbox{\tiny\tt 101}}}{w=5}\},\\
		&\{\stackrel{\mbox{\tiny\tt 101}}{v=5},\stackrel{\stackrel{\ \bullet\bullet}{\mbox{\tiny\tt 110}}}{w=6}\},
		\{\stackrel{\mbox{\tiny\tt 110}}{v=6},\stackrel{\stackrel{\ \ \bullet}{\mbox{\tiny\tt 111}}}{w=7}\},
		\{\stackrel{\mbox{\tiny\tt 111}}{v=7},\stackrel{\stackrel{\ \ \bullet}{\mbox{\tiny\tt 110}}}{w=6}\},
		\{\stackrel{\mbox{\tiny\tt 110}}{v=6},\stackrel{\stackrel{\ \bullet\bullet}{\mbox{\tiny\tt 101}}}{w=5}\},
		\{\stackrel{\mbox{\tiny\tt 101}}{v=5},\stackrel{\stackrel{\ \ \bullet}{\mbox{\tiny\tt 100}}}{w=4}\},\\
&{\color{greyout}\{\stackrel{\mbox{\tiny\tt 100}}{v=4},\stackrel{\stackrel{\bullet\bullet\bullet}{\mbox{\tiny\tt 011}}}{w=3}\},}
		\{\stackrel{\mbox{\tiny\tt 011}}{v=3},\stackrel{\stackrel{\ \ \bullet}{\mbox{\tiny\tt 010}}}{w=2}\},
		\{\stackrel{\mbox{\tiny\tt 010}}{v=2},\stackrel{\stackrel{\ \bullet\bullet}{\mbox{\tiny\tt 001}}}{w=1}\},
		\{\stackrel{\mbox{\tiny\tt 001}}{v=1},\stackrel{\stackrel{\ \ \bullet}{\mbox{\tiny\tt 000}}}{w=0}\}\\
		& {\bf (\yKstarLm=2,  \wordLength=3, k=2, m=1)}\\[2.0ex]
m=2 : \quad &\{\stackrel{\mbox{\tiny\tt 000}}{v=0},\stackrel{\stackrel{\ \bullet\ }{\mbox{\tiny\tt 010}}}{w=2}\},
		\{\stackrel{\mbox{\tiny\tt 001}}{v=1},\stackrel{\stackrel{\ \bullet\ }{\mbox{\tiny\tt 011}}}{w=3}\},
		\{\stackrel{\mbox{\tiny\tt 010}}{v=2},\stackrel{\stackrel{\bullet\bullet\ }{\mbox{\tiny\tt 100}}}{w=4}\},
		\{\stackrel{\mbox{\tiny\tt 011}}{v=3},\stackrel{\stackrel{\bullet\bullet\ }{\mbox{\tiny\tt 101}}}{w=5}\},
		\{\stackrel{\mbox{\tiny\tt 100}}{v=4},\stackrel{\stackrel{\ \bullet\ }{\mbox{\tiny\tt 110}}}{w=6}\},\\
		&\{\stackrel{\mbox{\tiny\tt 101}}{v=5},\stackrel{\stackrel{\ \bullet\ }{\mbox{\tiny\tt 111}}}{w=7}\},
		\{\stackrel{\mbox{\tiny\tt 110}}{v=6},\stackrel{\stackrel{\ \bullet\ }{\mbox{\tiny\tt 100}}}{w=4}\},
		\{\stackrel{\mbox{\tiny\tt 111}}{v=7},\stackrel{\stackrel{\ \bullet\ }{\mbox{\tiny\tt 100}}}{w=5}\},
		\{\stackrel{\mbox{\tiny\tt 101}}{v=5},\stackrel{\stackrel{\bullet\bullet\ }{\mbox{\tiny\tt 011}}}{w=3}\},
		\{\stackrel{\mbox{\tiny\tt 100}}{v=4},\stackrel{\stackrel{\bullet\bullet\ }{\mbox{\tiny\tt 010}}}{w=2}\},\\
		&\{\stackrel{\mbox{\tiny\tt 011}}{v=3},\stackrel{\stackrel{\ \bullet\ }{\mbox{\tiny\tt 001}}}{w=1}\},
		\{\stackrel{\mbox{\tiny\tt 010}}{v=2},\stackrel{\stackrel{\ \bullet\ }{\mbox{\tiny\tt 000}}}{w=0}\}\\
		& {\bf (\yKstarLm=2,  \wordLength=3, k=2, m=2)}\\[2.0ex]
m=3 : \quad &\{\stackrel{\mbox{\tiny\tt 000}}{v=0},\stackrel{\stackrel{\ \bullet\bullet }{\mbox{\tiny\tt 011}}}{w=3}\},
		\{\stackrel{\mbox{\tiny\tt 001}}{v=1},\stackrel{\stackrel{\bullet\ \bullet}{\mbox{\tiny\tt 100}}}{w=4}\},
{\color{greyout}\{\stackrel{\mbox{\tiny\tt 010}}{v=2},\stackrel{\stackrel{\bullet\bullet\bullet}{\mbox{\tiny\tt 101}}}{w=5}\},}
		\{\stackrel{\mbox{\tiny\tt 011}}{v=3},\stackrel{\stackrel{\bullet\ \bullet}{\mbox{\tiny\tt 110}}}{w=6}\},
		\{\stackrel{\mbox{\tiny\tt 100}}{v=4},\stackrel{\stackrel{\ \bullet\bullet}{\mbox{\tiny\tt 111}}}{w=7}\},\\
& {\color{greyout}\{\stackrel{\mbox{\tiny\tt 101}}{v=5},\stackrel{\stackrel{\bullet\bullet\bullet}{\mbox{\tiny\tt 010}}}{w=2}\},}
		\{\stackrel{\mbox{\tiny\tt 110}}{v=6},\stackrel{\stackrel{\bullet\ \bullet}{\mbox{\tiny\tt 011}}}{w=3}\},
		\{\stackrel{\mbox{\tiny\tt 111}}{v=7},\stackrel{\stackrel{\ \bullet\bullet}{\mbox{\tiny\tt 100}}}{w=4}\},
		\{\stackrel{\mbox{\tiny\tt 100}}{v=4},\stackrel{\stackrel{\bullet\ \bullet}{\mbox{\tiny\tt 001}}}{w=1}\},
		\{\stackrel{\mbox{\tiny\tt 011}}{v=3},\stackrel{\stackrel{\ \bullet\bullet}{\mbox{\tiny\tt 000}}}{w=0}\}\\
		& {\bf (\yKstarLm=2, \wordLength=3, k=2, m=3)}\\
\end{split}
\end{align*}
\iftoggle{pnasTemplate}{\vspace{-0.2in}}{}
\iftoggle{pnasTemplate}{\end{framed}}{}
\iftoggle{pnasTemplate}{\vspace{-0.2in}}{}
\caption{Enumeration of all possible cases of value deviation, $m
= \{1, 2, 3\}$, for $\wordLength=3$-bit words and a maximum of $k=2$
simultaneous upsets, for pairs of error-free values $v$ and
error-containing values $w$.  Pairs of values that are not possible
with a maximum of $k=2$ upsets are shown in a lighter color. Bit
positions which are in error are shown with a ``$\bullet$'' above
them. Note that, for $\wordLength=3$ and $k=2$, there is no combination
of three upsets that would lead to a value deviation of $m=2$. The
number of unique bit positions in which up to $k$ upsets can lead
to an integer distance of $m$ is given by the special function
$\yKstarLm$.}
\label{fig:L3k2mcases3}
\iftoggle{pnasTemplate}{\end{minipage}}{}
\iftoggle{pnasTemplate}{~~
\begin{minipage}[b]{0.25\textwidth}
\includegraphics[trim=0cm 0cm 0cm 0cm, clip=true, angle=0, width=1.0\textwidth]{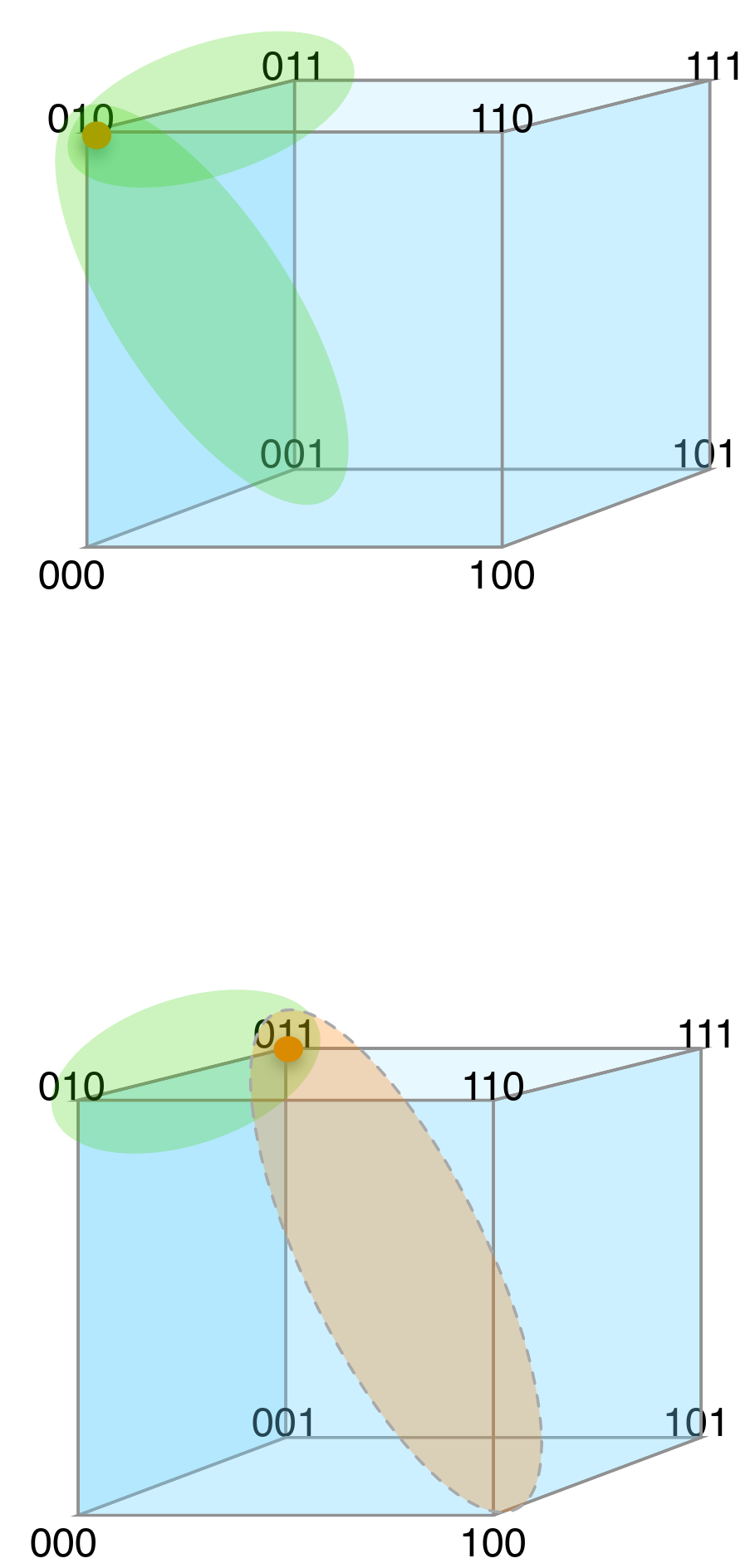}
\vspace{-0.2in}
\caption{For $\wordLength=3$ and integer distance $m=1$, pairs of
values such as $(v=2, w=3)$ and $(v=2, w=1)$ satisfy the constraint
of $k \le 2$ (top figure); on the other hand, while $(v=3, w=2)$
is a valid pair under the constraints, $(v=3, w=4)$ is not (bottom
figure).}
\vspace{-0.1in}
\label{fig:L3k2mcases3-geometric}
\end{minipage}
}{}
\end{figure*}

\begin{Definition}
\label{definition:yk*m*lkm}
The number of unique placements of less than or equal to $k$-bit
channel errors into an $\wordLength$-bit word, such that the resulting
change in value of the word is less than or equal to $m$ is given
by:
\begin{align*}
\yKstarLm=\left| \{x \in \Z_2^\wordLength : d_\Z^+(x, \hat{x})=m, d(x, \hat{x})\le k \} \right|.
\end{align*}
\end{Definition}
\noindent Figure~\ref{fig:L3k2mcases3} illustrates $\yKstarLm$  for $\wordLength=3$ and $k=2$.

\iftoggle{pnasTemplate}{\vspace{0.5in}}{}
\section{Tighter Bounds on Encoding Overhead}
\label{section:bounds}
To correct value erasures with the objective of minimizing integer
distortion, the syndrome of a forward error correcting code must
be able to distinguish between the ordered pairs of words
that differ in $k$ bit positions for some integer distortion $m$.
From prior work which defined bounds on encoding efficiency for VDB
codes~\cite{StanleyMarbell:itw09}, we have the following
Definition~\ref{definition:zLkm}.

\begin{Definition}
\label{definition:zLkm}
The number of ordered pairs of $\wordLength$-bit words that differ in $k$ bit
positions, and for which the absolute value of their integer
difference is exactly $m$ is given by:
\begin{align*}
\zLkm = \left| \{x \in \Z_2^l : d_{\Z^+}(x, \hat{x})=m, d(x, \hat{x})=k \} \right| .
\end{align*}
\end{Definition}

From~\cite{StanleyMarbell:itw09}, the function $\zLkm$ is the number
of solutions to the simultaneous Diophantine equation pair:
\begin{subeqnarray}
\label{eqn:numLkmSolutions:unsigned}
\left\vert\sum_{i = 0}^{\wordLength - 1}w_i 2^i - \sum_{i = 0}^{\wordLength - 1}v_i 2^i\right\vert = m ,\slabel{eqn:numLkmSolutions:unsigned:a}\\
\sum_{i = 0}^{\wordLength - 1}\left(w_i(1-v_i) + v_i(1-w_i) \right) = k \slabel{eqn:numLkmSolutions:unsigned:b}.
\end{subeqnarray}

From~\cite{StanleyMarbell:itw09}, an upper bound\footnote{The bound
on $\zKstarLm$ given in ~\cite{StanleyMarbell:itw09} is missing a
factor of $k$ since $\zKstarLm$ should pessimistically sum $\zLkm$
over all cases of number of upsets less than or equal to $k$.} on
the number of solutions to Equation~\ref{eqn:numLkmSolutions:unsigned}
is
\begin{align}
\label{eqn:zbound}
\zLkm	\le	 2^{\wordLength+1} - 2m .
\end{align}
From the results of numerical simulation, we observe that the number of solutions is
always 0, 1, or a multiple of $2^{\wordLength-k+1}$. Therefore, a tighter bound is 
\begin{align}
\label{eqn:tighterzLkmBound}
\zLkm \le	 2^{\wordLength+1} - 2m - \mod(2^{\wordLength+1} - 2m, 2^{\wordLength-k+1}) .
\end{align}

\begin{figure}[t]
\centering
\subfloat[Bounds for $\wordLength=8, k=3$.]{\includegraphics[trim=0cm 0cm 0cm 0cm, clip=true, angle=0, width=0.245\textwidth]{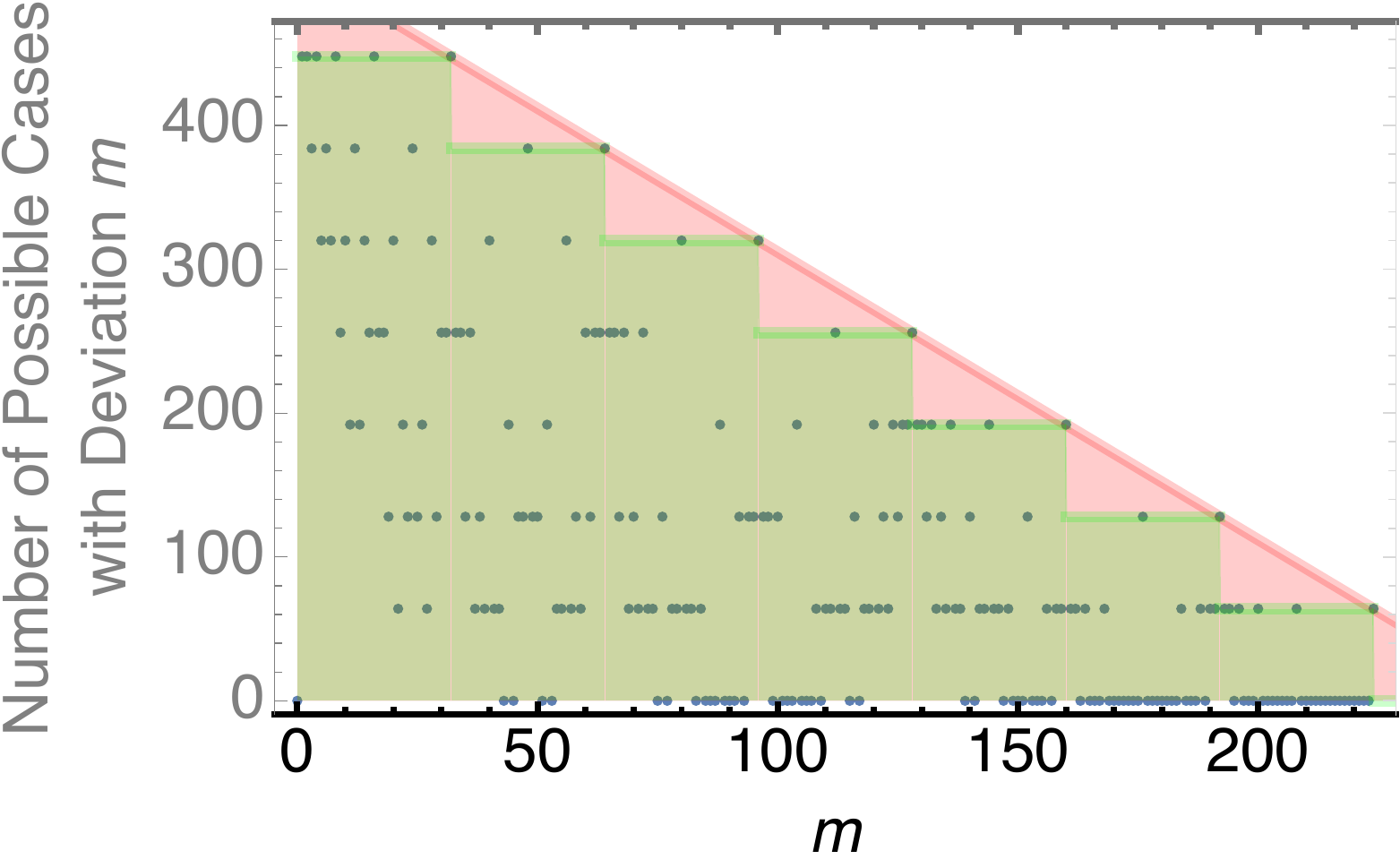}}~
\subfloat[Bounds for $\wordLength=8, k=4$.]{\includegraphics[trim=0cm 0cm 0cm 0cm, clip=true, angle=0, width=0.245\textwidth]{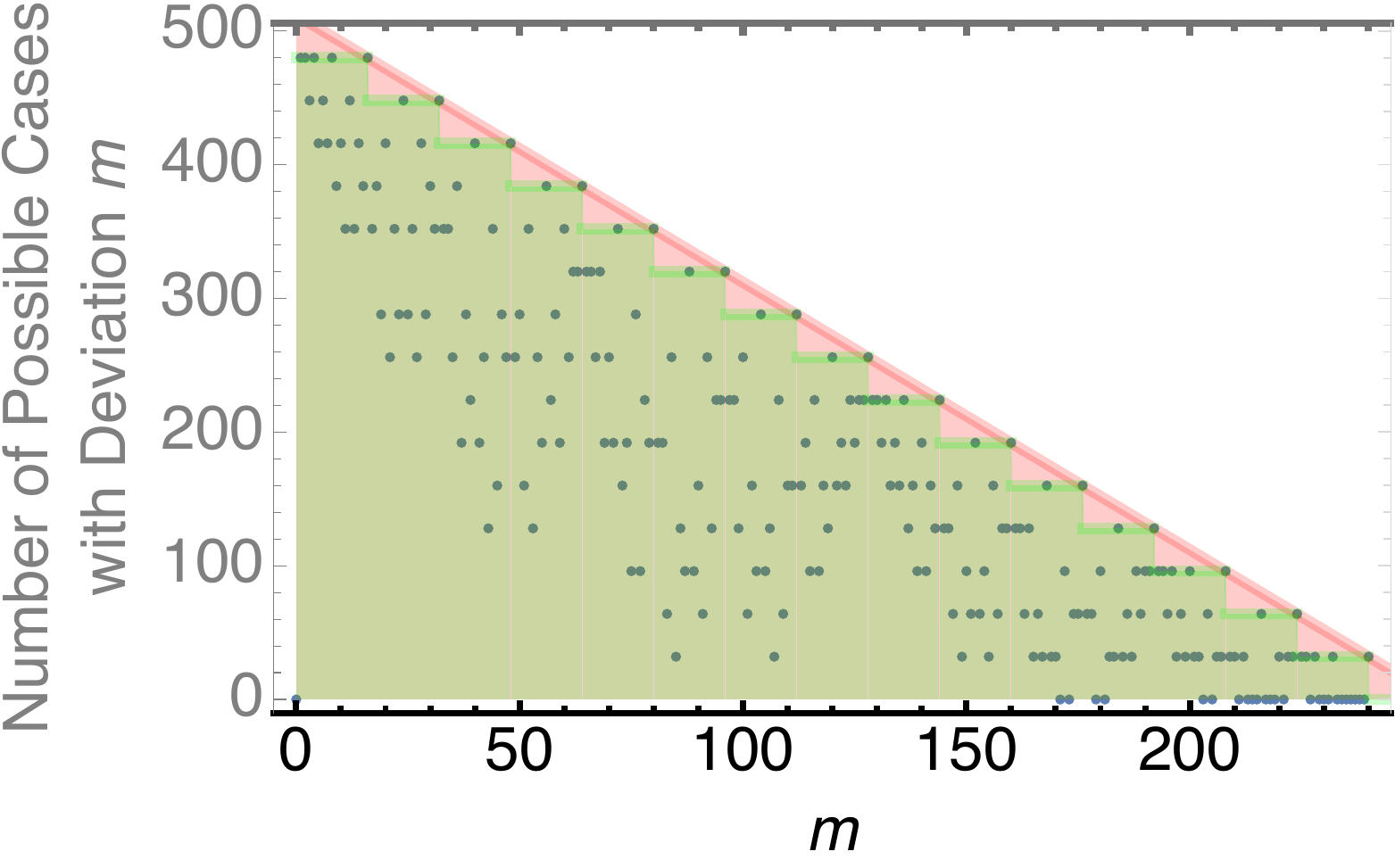}}
\vspace{-0.1in}
\caption{The tighter bounds we introduce in
Equation~\ref{eqn:tighterzLkmBound} (staircase, green shaded region),
compared to the previous best bound from~\cite{StanleyMarbell:itw09}
(Equation~\ref{eqn:zbound}, triangular red shaded region). The
bounds of Equation~\ref{eqn:tighterzLkmBound} are the tightest
possible monotonic bounds on $\zLkm$.}
\vspace{-0.1in}
\label{fig:zlkmBoundImproved}
\end{figure}

\subsection{Empirical evaluation of tighter bounds}
\label{section:tighertBoundsEvaluation}
Figure~\ref{fig:zlkmBoundImproved} shows, by numerical simulation, how the bound
given in Equation~\ref{eqn:tighterzLkmBound} for the quantity $\zLkm$
from Definition~\ref{definition:zLkm} compares to the previous best
bound from~\cite{StanleyMarbell:itw09} (Equation~\ref{eqn:zbound}).
The numerical simulation results show that the bound we present in
Equation~\ref{eqn:tighterzLkmBound} is the tightest possible monotonic
bound.

\section{\iftoggle{pnasTemplate}{Probabilistic Value Deviation Bounded Encoding}{Probabilistic VDB Encoding}}
\label{section:encoding}
We define a {\it constraint tail distribution}
$\widehat{\mathrm{F}}_{\!M}(m)$ to denote the distribution of integer
distances $d_{\Z^+}(x, \hat{x})$ which an algorithm or application
consuming the output of a communication channel can tolerate.  Its domain is
given by Equation~\ref{eqn:maxm} and it may be any monotone
non-decreasing function of $m$ with range $[0,1]$.  We first ignore
the distribution of values transmitted on the channel and consider
the cases in which no channel errors are masked
(Section~\iftoggle{pnasTemplate}{3\,}{}\ref{section:iidBernoulli} and
Section~\iftoggle{pnasTemplate}{3\,}{}\ref{section:nonIidErrors}).
\iftoggle{pnasTemplate}{
We introduce the effect of the
distribution of channel values in Section~3\,\ref{section:maskedErrors}.}{}

\subsection{Case 1: i.i.d. Bernoulli channel errors}
\label{section:iidBernoulli}
If channel errors are independent and identically distributed (i.i.d.)
Bernoulli events with parameter $p$, then $k$
channel errors in an $\wordLength$-bit word will be induced by the channel with
probability
\begin{align}
	\binom{\wordLength}{k} p^k(1-p)^{\wordLength-k} .
\end{align}
Not all $\binom{\wordLength}{k}$ placements of $k$ errors in $\wordLength$-bit words
leading to integer distortion $m$ are possible.  The subset that
is possible is given by $\yKstarLm$ of
Definition~\ref{definition:yk*m*lkm}, the number of unique ways in
which the channel can induce up to $k$ errors into an $\wordLength$-bit word
$x$, such that the resulting distance $d_{\Z^+}(x, \hat{x})$ of the
channel output $\hat{x}$ from the channel input $x$, is exactly
$m$.  The events constituting the placements of errors are disjoint
and thus the probability of {\it any} placement of up to $k$ channel
errors in an $\wordLength$-bit word leading to a deviation of $m$ is given
by the sum of their (identical) individual probabilities.

Let $\mathbb{S}_{\langle \yKstarLm \rangle}$ be the set of sets of
$\wordLength$-bit vectors, one for each induced distance $m$, with \logicone
in a vector indicating a bit position in which the up to $k$ bit-level
disturbances must occur in order to lead to a value deviation $m$
and  \logiczero otherwise.  For example, for $\wordLength=3$ and
$k=2$, $\{$\logiczero\logiczero\logicone, \logiczero\logicone\logicone$\}$
is the member of $\mathbb{S}_{\langle \yKstarLm \rangle}$ corresponding
to the distance $m=1$.  These bit vectors correspond to the positions
of the black filled circles in Figure~\ref{fig:L3k2mcases3}. Let
the $i$'th bit position of vector $e \in \mathbb{S}_{\langle \yKstarLm
\rangle}$ be $e_{\langle i\rangle}$.  For the constraint of the
tail distribution on tolerable deviations to be satisfied,
\begin{align}
\sum_{e \in \mathbb{S}_{\langle \yKstarLm \rangle}}\prod_{i=0}^{\wordLength-1} p^{e_{\langle i\rangle}}(1 - p)^{1 - e_{\langle i\rangle}} \le \widehat{\mathrm{F}}_{\!M}(m), ~~ \forall m .\label{eqn:constraintIidBernoulli}
\end{align}
Thus, given the constraint $\widehat{\mathrm{F}}_{\!M}(m)$ on the
tolerable distribution of distances induced by the channel, we can determine the
maximum tolerable channel error probability, $p$.

If larger tolerable channel error probabilities lead to better
resource savings, then the most resource-efficient encoding will
be that which assigns the largest value to $p$ while still
satisfying the constraint $\widehat{\mathrm{F}}_{\!M}(m)$.  We can obtain the value of $p$ by
solving the optimization problem
\begin{align}
\label{eqn:minimizationIidBernoulli}
& \underset{}{\text{maximize}}
& & p \\\nonumber
& \text{subject to}
& & \sum_{e \in \mathbb{S}_{\langle \yKstarLm \rangle}}\prod_{i=0}^{\wordLength-1} p^{e_{\langle i\rangle}}(1 - p)^{1 - e_{\langle i\rangle}} & \le \widehat{\mathrm{F}}_{\!M}(m) .
\end{align}
For a given $\widehat{\mathrm{F}}_{\!M}(m)$, we solve the corresponding
linear program for $p$ in Equation~\ref{eqn:minimizationIidBernoulli}.
Channel errors may be masked and the value of $p$ denotes the
probability of channel errors after considering the effects of
masking. \iftoggle{pnasTemplate}{We address masking in Section~3\,\ref{section:maskedErrors}.}{}

\iftoggle{pnasTemplate}
{\begin{figure}
\centering
{\includegraphics[trim=0cm 0cm 0cm 0cm, clip=true, angle=0, width=0.35\textwidth]{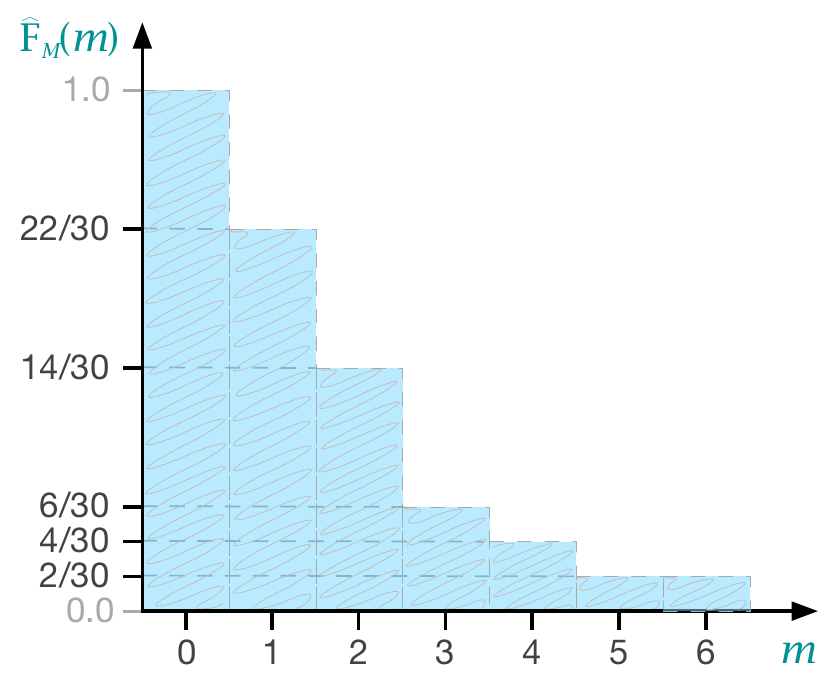}}
\vspace{-0.1in}
\caption{An example constraint distribution
$\widehat{\mathrm{F}}_{\!M}(m)$, determining the probability
distribution of integer distances of transmitted words.}
\vspace{-0.1in}
\label{fig:FmHat}
\end{figure}
}
{}

\paragraph{Example: $\wordLength=3$ and $k=2$}
\label{section:iidBernoulliExample}
Consider a communication channel in which the transmitted words are
$\wordLength=3$ bits long and where we can stochastically induce at most $k=2$
bit errors in a way that lowers the power dissipated by the channel's
transmitter.  From Equation~\ref{eqn:maxm}, when $\wordLength=3$, $k=2$ and
for a binary symmetric channel, the range of possible integer
distortion values is $1 \le m \le 6$. The set $\mathbb{S}_{\langle
\yKstarLm \rangle}$ is
$\langle$
$\{$\logiczero\logiczero\logicone, \logiczero\logicone\logicone$\} : m=1$, 
$\{$\logiczero\logicone\logiczero, \logicone\logicone\logiczero$\} : m=2$,
$\{$\logiczero\logicone\logicone, \logicone\logiczero\logicone$\} : m=3$,
$\{$\logicone\logiczero\logiczero$\} : m=4$,
$\{$\logicone\logiczero\logicone$\} : m=5$, 
$\{$\logicone\logicone\logiczero$\} : m=6$ $\rangle$
with cardinalities (i.e., corresponding $\yKstarLm$) $\langle 2, 2, 2, 1, 1, 1\rangle$.
For the individual cases of $m$, Equation~\ref{eqn:minimizationIidBernoulli} becomes
\begin{align}
\label{eqn:iidBernoulliExample}
& \underset{}{\text{maximize}} & p \\\nonumber
& \text{subject to}\\\nonumber
			& & ~~ p(1 - p)^{2} + p^{2}(1 - p) & \le \widehat{\mathrm{F}}_{\!M}(1)	~~ \wedge\\[-0.9ex]\nonumber
	 		& & ~~ p(1 - p)^{2} + p^{2}(1 - p) & \le \widehat{\mathrm{F}}_{\!M}(2)	~~ \wedge\\\nonumber
			& & ~~ p(1 - p)^{2} + p^{2}(1 - p) & \le \widehat{\mathrm{F}}_{\!M}(3)	~~ \wedge\\\nonumber
			& & ~~ p(1 - p)^{2}		& \le \widehat{\mathrm{F}}_{\!M}(4)	~~ \wedge\\\nonumber
			& & ~~ p^{2}(1 - p)		& \le \widehat{\mathrm{F}}_{\!M}(5)	~~ \wedge\\\nonumber
			& & ~~ p^{2}(1 - p)		& \le \widehat{\mathrm{F}}_{\!M}(6)	.
\end{align}

For example,
\iftoggle{pnasTemplate}
{if the deviations occur with the tail distribution
shown in Figure~\ref{fig:FmHat},}
{choosing for the sake of illustration $\widehat{\mathrm{F}}_{\!M}(0) = 1$, $\widehat{\mathrm{F}}_{\!M}(1) = \sfrac{22}{30}$, $\widehat{\mathrm{F}}_{\!M}(2) = \sfrac{14}{30}$, $\widehat{\mathrm{F}}_{\!M}(3) = \sfrac{6}{30}$, $\widehat{\mathrm{F}}_{\!M}(4) = \sfrac{4}{30}$, $\widehat{\mathrm{F}}_{\!M}(5) = \sfrac{2}{30}$, and $\widehat{\mathrm{F}}_{\!M}(6) = \sfrac{2}{30}$,}
then
Equation~\ref{eqn:iidBernoulliExample} becomes
\begin{align}
& \underset{}{\text{maximize}} & p \\\nonumber
& \text{subject to}\\\nonumber
			& & ~~ p(1 - p)^{2} + p^{2}(1 - p) & \le \sfrac{6}{30}	~~ \wedge\\[-0.9ex]\nonumber
			& & ~~ p(1 - p)^{2}		 & \le \sfrac{4}{30}	~~ \wedge\\\nonumber
			& & ~~ p^{2}(1 - p)		 & \le \sfrac{2}{30}	.
\end{align}
This yields $p=0.2180$ for all $m$. We describe how we automate the 
solution of $p$ in Section~\ref{section:howWeSolveUsingDreal}.

We can exploit this constraint on bit-level error in one of two
ways. First, in a hardware implementation where communications are
otherwise error-free, up to $k=2$ bits could be purposefully
transmitted in error with probability $p=0.2180$ if doing so would
reduce communication power dissipation. Second, we may exploit the constraint
on $p$ in hardware if we can change properties of the channel.
Examples of such changes include the operation of the link at
lower-than-nominal voltages, or the removal of link termination,
both of which increase bit error rates, but yield a potential saving
in energy per bit.

Under these assumptions of i.i.d. Bernoulli channel errors however,
we cannot discriminate between which bits the energy saving
techniques are applied to. To address this weakness, we next consider a
channel model where the channel can induce different error probabilities
across the ordinal positions of the bits of a word transmitted over
the channel.

\subsection{Case 2: Independent non-identically-distributed errors}
\label{section:nonIidErrors}
When the probabilities $p_i$ of channel errors that may be induced at different bit positions $i$
in a transmitted word are not identical, the constraint on error
probabilities is a generalization of
Equation~\ref{eqn:constraintIidBernoulli}:
\begin{align}
\sum_{e \in \mathbb{S}_{\langle \yKstarLm \rangle}}\prod_{i=0}^{\wordLength-1} p_i^{e_{\langle i\rangle}}(1 - p_i)^{1 - e_{\langle i\rangle}} \le \widehat{\mathrm{F}}_{\!M}(m), ~~ \forall m .
\end{align}

Again, if the channel can reduce resource usage of a transmitter on the channel by permitting larger error
probabilities $p_i$, then the most efficient encoding is that which
assigns the largest value to each $p_i$ while still satisfying the
constraint. We can obtain the probabilities by solving the optimization problem
\begin{align}
\label{eqn:minimizationNonIidBernoulli}
& \underset{}{\text{maximize}}
& & p_i \\\nonumber
& \text{subject to}\\\nonumber
			& & \sum_{e \in \mathbb{S}_{\langle \yKstarLm \rangle}}\prod_{i=0}^{\wordLength-1} p_i^{e_{\langle i\rangle}}(1 - p_i)^{1 - e_{\langle i\rangle}} & \le \widehat{\mathrm{F}}_{\!M}(m) .	
\end{align}
For a given $\widehat{\mathrm{F}}_{\!M}(m)$, we solve the corresponding
geometric program for $p_i$.  Channel errors may be masked and the
$p_i$ denote the probability of channel errors after considering
the effects of masking.

\paragraph{Example: $\wordLength=3$, $k=2$}
From Equation~\ref{eqn:maxm}, when $\wordLength=3$, $k=2$ and for a
binary symmetric channel, $1 \le m \le 6$. The possible values of
$\yKstarLm$ for each of these values of $m$ and the set
$\mathbb{S}_{\langle \yKstarLm \rangle}$ are the same as in
Section~\ref{section:iidBernoulliExample}. Considering the same
constraint $\widehat{\mathrm{F}}_{\!M}(m)$ on induced distance as
in Section~\ref{section:iidBernoulliExample},
then Equation~\ref{eqn:minimizationNonIidBernoulli} becomes
\begin{align}
\label{eqn:NonIidBernoulliExample}
& \underset{}{\text{maximize}} & p_i \\\nonumber
& \text{subject to}\\\nonumber
			& & ~~ p_0 - p_2 p_0			& \le \sfrac{22}{30}	~~ \wedge\\[-0.9ex]\nonumber
			& & ~~ p_1 - p_1 p_0			& \le \sfrac{14}{30}	~~ \wedge\\\nonumber
			& & ~~ (p_2 + p_1 - 2 p_2 p_1) p_0	& \le \sfrac{6}{30}	~~ \wedge\\\nonumber
			& & ~~ p_2 (p_1 - 1) (p_0 - 1)		& \le \sfrac{4}{30}	~~ \wedge\\\nonumber
			& & ~~ p_2 (1 - p_1) p_0		& \le \sfrac{2}{30}	~~ \wedge\\\nonumber
			& & ~~ p_2 p_1 (1 - p_0)		& \le \sfrac{2}{30}	.
\end{align}
This yields $p_2=0.2266$, $p_1=0.4011$, and $p_0 = 0.4485$ for all
$m$. We describe how we automate this solution with a software tool,
in Section~\ref{section:howWeSolveUsingDreal}. Given any constraint
$\widehat{\mathrm{F}}_{\!M}(m)$, we can compute the corresponding
$p_i$s for any given word length $\wordLength$ and maximum number
of simultaneous channel errors, $k$.

\subsection{How we solve for $p$ and $p_i$}
\label{section:howWeSolveUsingDreal}
To solve for $p$ in Equation~\ref{eqn:minimizationIidBernoulli} and
$p_i$ in Equation~\ref{eqn:minimizationNonIidBernoulli}, as we do
for the examples in Equation~\ref{eqn:iidBernoulliExample} and
Equation~\ref{eqn:NonIidBernoulliExample} we 
developed a software tool to compile
Equation~\ref{eqn:minimizationIidBernoulli} and
Equation~\ref{eqn:minimizationNonIidBernoulli} to constraints in
the SMT-2 format. We then use the dReal SMT solver~\cite{gao2013dreal} to solve for $p$
in Equation~\ref{eqn:iidBernoulliExample} or $p_i$ in
Equation~\ref{eqn:NonIidBernoulliExample}.

\iftoggle{pnasTemplate}{

\begin{figure}[t]
\centering
\includegraphics[trim=0cm 0cm 0cm 0cm, clip=true, angle=0, width=0.16\textwidth]{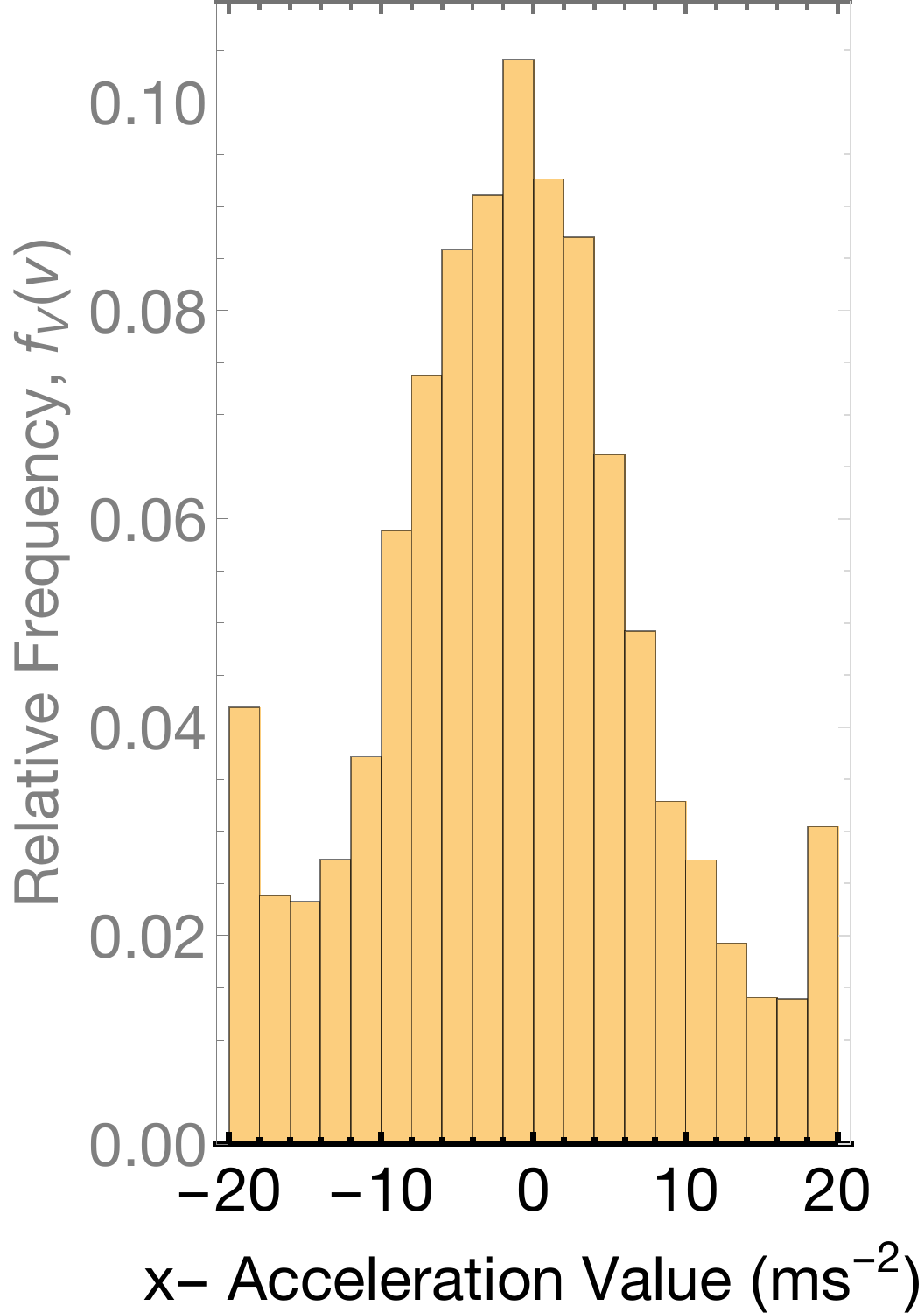}~
\includegraphics[trim=0cm 0cm 0cm 0cm, clip=true, angle=0, width=0.16\textwidth]{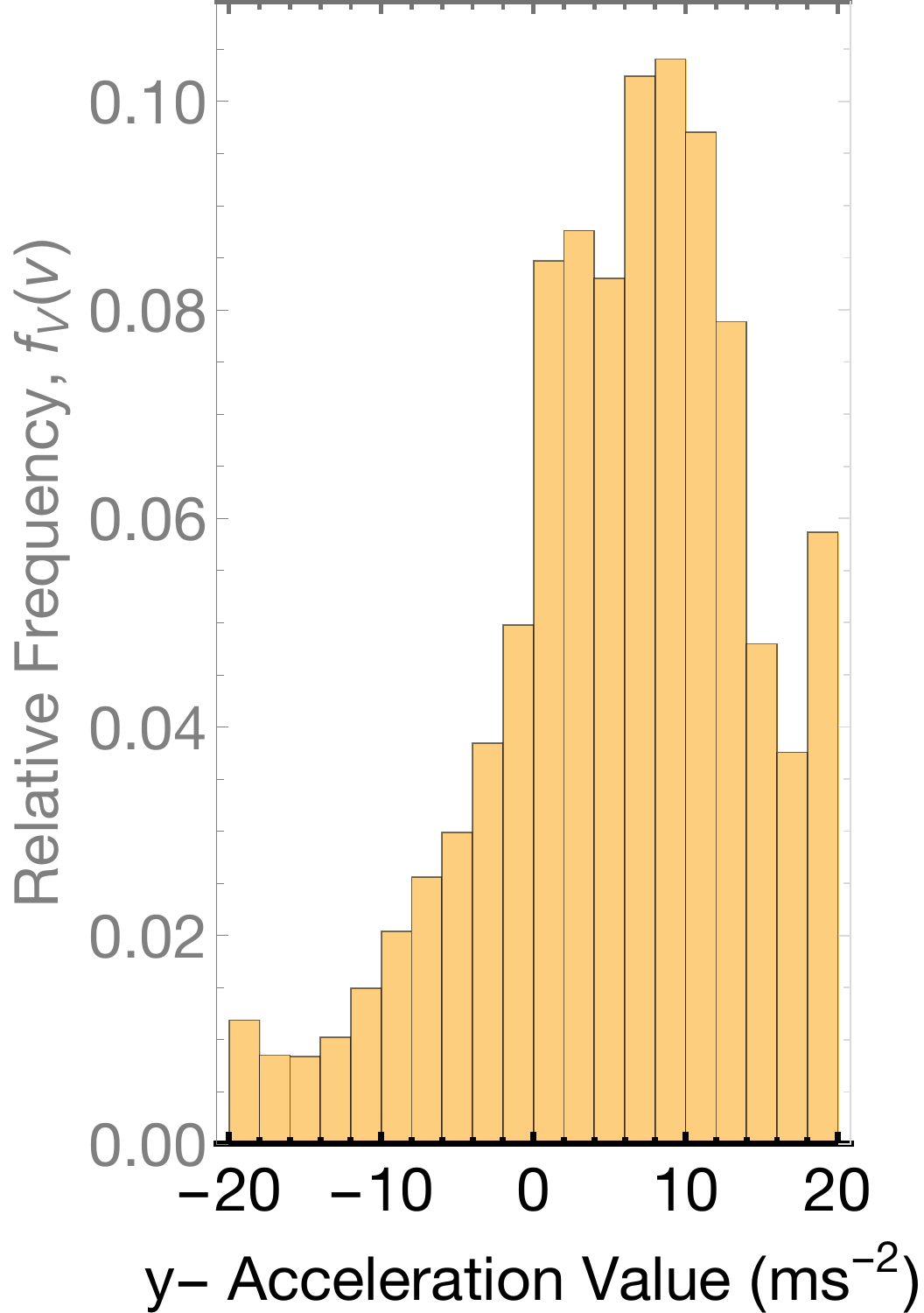}~
\includegraphics[trim=0cm 0cm 0cm 0cm, clip=true, angle=0, width=0.16\textwidth]{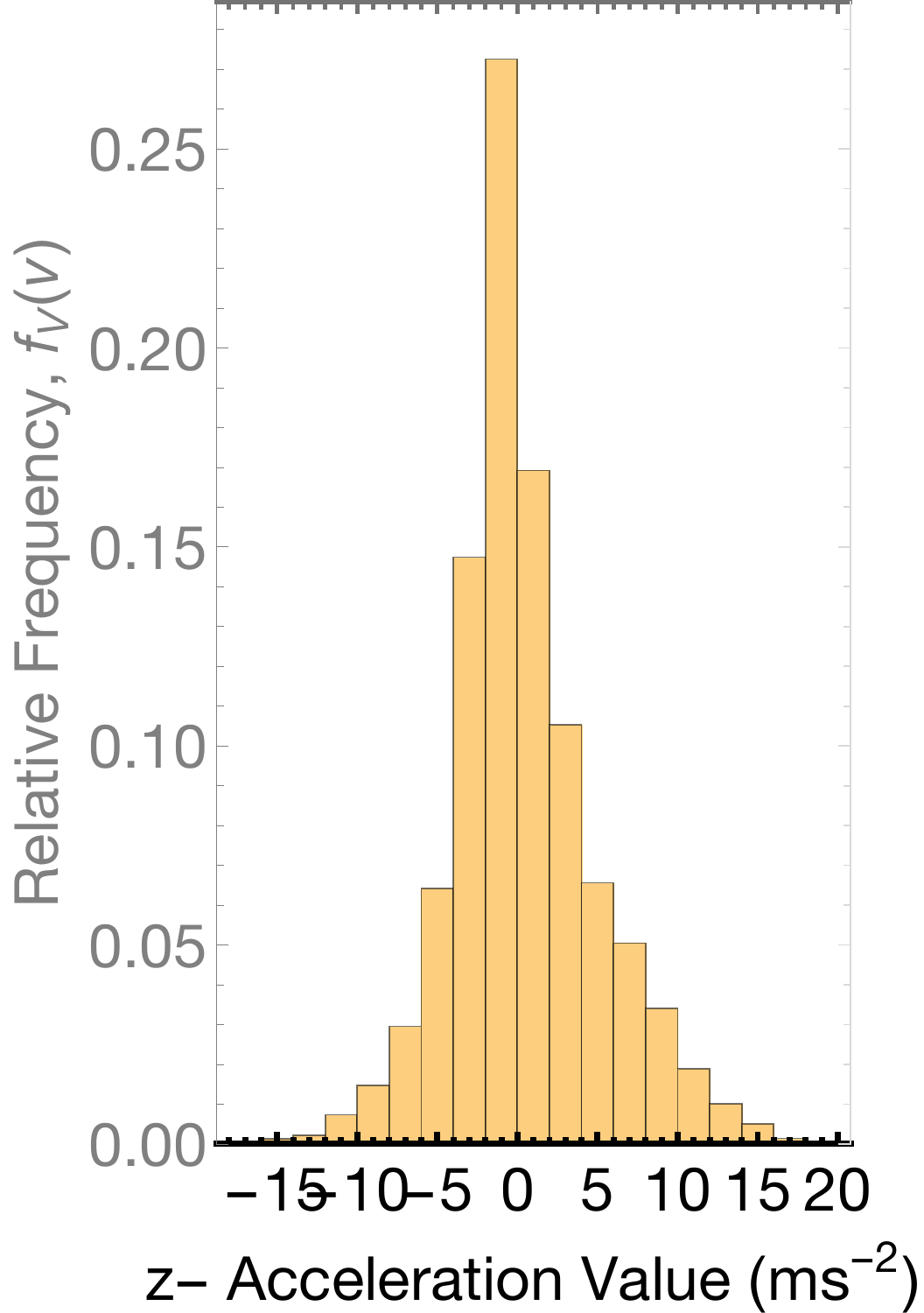}
\vspace{-0.1in}
\caption{Empirical probability mass functions (i.e., $f_{\!V}(v)$
from Figure~\ref{fig:fmeqns}) for 29\,978 samples (25 minutes of
sensor data sampled at 20\,Hz) from the three axes of an accelerometer
worn by a person engaged in a variety of
activities~\protect\cite{Kwapisz:2011}.}
\vspace{-0.1in}
\label{fig:WISDM_ar_v1.1-user1-fv-distributions}
\end{figure}

\begin{figure*}
\begin{framed}
\small
\begin{align}
\begin{split}
f_M(m)		&= \sum_{a=0}^{2^n - 1} \Biggl( \sum_{i=0}^{n-1}f_{\!V}(a+2^i)f_{\!t\langle i \rangle}(0)
		+ \sum_{i=0}^{n-1}f_{\!V}(a-2^i)f_{\!t\langle i \rangle}(1)
		+ \sum_{i=0}^{n-1}f_{\!t\langle i \rangle}(V_{\!\langle i \rangle})f_{\!V}(a)\Biggr) f_{\!V}(a-m)
		+ \sum_{a=0}^{2^n - 1} \Biggl( \sum_{i=0}^{n-1}f_{\!V}(a+2^i)f_{\!t\langle i \rangle}(0)\\
		&+ \sum_{i=0}^{n-1}f_{\!V}(a-2^i)f_{\!t\langle i \rangle}(1)
		+ \sum_{i=0}^{n-1}f_{\!t\langle i \rangle}(V_{\!\langle i \rangle})f_{\!V}(a)\Biggr) f_{\!V}(a+m) .\label{eqn:fmfromvwsie}
\end{split}
\end{align}
\begin{align}
\begin{split}
f_{\!M}(m) 	&= \sum_{b=0}^{2^n - 1} \Biggl(\sum_{a=0}^{2^n-1}\frac{\Pr\left \{  V=a, \sum_{i=0}^{n-1} M_i=\mid b-a\mid   \right \}}{f_{\!V}(a)} \Biggr) f_{\!V}(b-m)
		+ \sum_{b=0}^{2^n - 1} \Biggl(\sum_{a=0}^{2^n-1}\frac{\Pr\left \{  V=a, \sum_{i=0}^{n-1} M_i=\mid b-a\mid   \right \}}{f_{\!V}(a)} \Biggr) f_{\!V}(b+m), \label{eqn:fmfromvwmie}\\
&\mbox{where\quad} M_i	= \left( -\left( a_{\!\langle i \rangle}2^i\right) + \left(t_{\!\langle i \rangle}2^i \right) \right) \left( t_{\!\langle i \rangle} \oplus a_{\!\langle i \rangle}\right) .
\end{split}
\end{align}
\normalsize
\end{framed}
\caption{Analytic relations between the probability distribution of
error-free values ($f_{\!V}(v)$), the distribution of channel errors
($f_{\!t\langle i \rangle}(k)$), and the distribution of value
deviations ($f_{\!M}(m)$), from~\cite{stanley2008deviation}, for
the cases of a singly-occurring error (i.e., $k = 1$) assumption
(Equation~\ref{eqn:fmfromvwsie}) and a multiply-occurring error (i.e., $k \ge 1$)
assumption (Equation~\ref{eqn:fmfromvwmie}).}
\label{fig:fmeqns}
\end{figure*}

\subsection{Case 3: Accounting for distributions of values transmitted}
\label{section:maskedErrors}
The analyses in Section~\ref{section:iidBernoulli} and
Section~\ref{section:nonIidErrors} do not account for the distribution
of values transmitted over the channel and therefore do not account
for the fact that channel errors may be masked.

Let $f_{\!V}(v)$ (equivalently, $\Pr\left\{V=v\right\}$) be the
probability mass function for values transmitted on the channel,
let $f_{\!t\langle i \rangle}$ be the probability of bit error
induced by a VDB channel at bit position $i$, and let $f_{\!M}(m)$
be the probability mass function for the channel integer distortion.
Figure~\ref{fig:WISDM_ar_v1.1-user1-fv-distributions} shows real-world empirical examples of
the probability mass function $f_{\!V}(v)$ for three sets of data taken from
a real-world sensor deployment~\cite{Kwapisz:2011}.

Figure~\ref{fig:fmeqns} shows the analytical formulation for $f_{\!M}(m)$ when the channel induces
single bit errors (Equation~\ref{eqn:fmfromvwsie}) and multiple bit
errors (Equation~\ref{eqn:fmfromvwmie}). Our goal is to obtain the
function $f_{\!t\langle i \rangle}$ which satisfies a given constraint
$\widehat{\mathrm{F}}_{\!M}(m)$ on channel integer distortion.

For a given $\widehat{\mathrm{F}}_{\!M}(m)$, we obtain $f_{\!t\langle
i \rangle}$ from the definition of the tail distribution of probability
mass functions.  Let $m_{\mathrm{max}}$ be the maximum integer
distortion possible on the channel, and given previously in
Equation~\ref{eqn:maxm}. Then,
\begin{align}
	\sum\limits_{0}^{m_{\mathrm{max}}} f_{\!M}(m) = \widehat{\mathrm{F}}_{\!M}(m) .
\end{align} 
}{}

\section{The $\mathbb{S}_{\langle \yKstarLm \rangle}$ Sets\iftoggle{pnasTemplate}{ Required for VDB Code Tables}{}}
\label{section:implementation}
The encoding (i.e., determination of $p$s or $p_i$s) in
Section~\ref{section:encoding} depends on having the relevant set
$\mathbb{S}_{\langle \yKstarLm \rangle}$ for a given word length
$\wordLength$, maximum number of channel errors $k$,  and integer
distortion $m$.

\subsection{Basic construction of the sets $\mathbb{S}_{\langle \yKstarLm \rangle}$}
One pragmatic method to obtain $\mathbb{S}_{\langle \yKstarLm \rangle}$,
when dealing with small word sizes (e.g., $\wordLength$ of 8--16)
is to simply enumerate all the \iftoggle{pnasTemplate}{solution $\langle w, v\rangle$ pairs}{solutions}
to \iftoggle{pnasTemplate}{Equation~\ref{eqn:numLkmSolutions:unsigned}}{the equations defining pairs of values that have Hamming distance $k$ and integer distance $m$} and to determine the
bit positions in which they differ.

\subsection{Efficient construction of the sets $\mathbb{S}_{\langle \yKstarLm \rangle}$}
Let $\integerDistance$ be an integer distance. We can write
$\integerDistance$ in any radix, $\encodingRadix$, as a polynomial
$\sum_i \beta_i \encodingRadix^i$. Let $\errorFreeValue(\encodingRadix)$
be the radix-$\encodingRadix$ polynomial representation of a value
$\errorFreeValue$, with the coefficients $\errorFreeValue_i$ of the
radix-$\encodingRadix$ expansion, i.e.,
$\errorFreeValue(\encodingRadix)=\sum_i \errorFreeValue_i
\encodingRadix^i$. Let $\errorContainingValueSet_\integerDistance$
be the set of all polynomials
$\errorFreeValue(\encodingRadix)(\encodingRadix - 2) +
\integerDistance(\encodingRadix)$ which have no terms with coefficients
$\pm 2$, that is
\begin{align}
	\beta_0 - 2\errorFreeValue_0 \ne \pm 2\\
	\beta_i - 2\errorFreeValue_i + \errorFreeValue_{i-1} \ne \pm 2 .
\end{align}
Then, given $\integerDistance$, and a value $s$,
Algorithm~\ref{alg:BranchingErrorWords} computes the members of the
set $\errorContainingValueSet_\integerDistance$ of values within
integer distance $\integerDistance$ of the value $\errorFreeValue$.
Given this set $\errorContainingValueSet_\integerDistance$, we then
extract the subset which are at Hamming distance $k$ or smaller and
list the bit positions in which the elements of that subset differ
from $s$, to get the set $\mathbb{S}_{\langle \yKstarLm \rangle}$.

\begin{algorithm}
\footnotesize
\DontPrintSemicolon
\SetKwProg{Proc}{Procedure}{}{end}
\KwData{An integer deviation $\integerDistance$ with coefficients
$\beta_i$ in its radix-$\encodingRadix$ expansion, i.e.,
$\integerDistance(\encodingRadix)=\sum_i \beta_i \encodingRadix^i$
and an integer $\errorFreeValue$ with coefficients $\errorFreeValue_i$
of its radix-$\encodingRadix$ expansion, i.e., 
$\errorFreeValue(\encodingRadix)=\sum_i \errorFreeValue_i \encodingRadix^i$,
and a word length $L$.
}
\KwResult{The set $\errorContainingValueSet_\integerDistance$ of all 
$L$-bit values that are at integer distance $m$ from $s$: These have polynomial expansions of the form $\polynomial(\encodingRadix) = \errorFreeValue(\encodingRadix)(\encodingRadix - 2) + m(\encodingRadix)$
having only coefficients $0,\pm 1$, and where $\polynomial(2)=m$.}

\Proc{CalculateSequences ($m,L$)}
{
	\tcc{{\footnotesize Compute the radix-$\encodingRadix$ expansion of $m$:}}	
	$\beta \leftarrow$ radix-$\encodingRadix$ expansion $m(\encodingRadix)= \sum_i \beta_i \encodingRadix^i$ from $m$\;	

	\tcc{{\footnotesize Recursively compute all polynomials $s(\encodingRadix)$:}}	
	$V_m \leftarrow$ Branch($0, b, s, 0, L$)\;

	\tcc{{\footnotesize Put the result together with $m(\encodingRadix)$:}}	
	$S_m \leftarrow \{ v(\encodingRadix)(\encodingRadix-2)+ m(\encodingRadix) : v(\encodingRadix) \in V_m\}$\;

	\KwRet{$S_m$}
}

\Proc{CheckAndBranch($w, \beta, t, i, L$)}
{
	\eIf{$ i = L-1$}
	{
		\tcc{{\footnotesize Cannot go any further.}}	
		\KwRet{$t$}\;
	}
	{
		\tcc{{\footnotesize Keep going with new branch:}}	
		\KwRet{Branch($w, \beta, t, i+1, L$)}\;
	}
}

\Proc{Branch($w, \beta, s, i, L$)}
{
	$S \leftarrow \emptyset$\;
	\While{$\beta_i = w$}
	{
		\tcc{{\footnotesize Go until $\beta_i$ changes or word end.}}	
		$s_i \leftarrow w$\;
		$i \mathrel{+}= 1$\;
		\If{$ i = L$}
		{
			\tcc{{\footnotesize Reached end of word.}}	
			$S \leftarrow S \bigcup s$\;
			\textbf{return}  $S$\;
		}
	}
	
	$t \leftarrow s$\;
	$t_i  \leftarrow 0$\;
	$S \leftarrow S \bigcup$ CheckAndBranch($0, \beta, t, i, L$)\;
	
	\tcc{{\footnotesize New branch: starts with $1$ in the $i$th place:}}	
	$t \leftarrow s$\;
	$t_i  \leftarrow 1$\;
	$S \leftarrow S \bigcup$ CheckAndBranch($1, \beta, t, i, L$)\;
	\KwRet{$S$}
}

\label{alg:BranchingErrorWords:H}
\normalsize
\caption{Algorithm for finding all values within Hamming distance $k$ and integer distance $m$ from a value $s$.}
\label{alg:BranchingErrorWords}
\end{algorithm}

\section{Validating Encoding\iftoggle{pnasTemplate}{ using Monte Carlo Simulation}{}}
\label{section:monteCarlo}
To demonstrate the effectiveness of probabilistic VDB encoding, we
compute the code tables (i.e., the tolerable error at each ordinal
bit position in a word), given a constraint on the tail distribution
of integer distortion (i.e., $\widehat{\mathrm{F}}_{\!M}(m)$).  Then, given
the probability with which each bit can be in error (i.e., the code
table), we perform a Monte Carlo simulation, inducing bit
errors with the rates dictated by the code table. We compare the
distribution of the resulting integer distortion with the constraint
on tail distribution of integer distortion $\widehat{\mathrm{F}}_{\!M}(m)$.
We use a word size $L = 3$, a maximum number of channel errors
$k = 3$, and a constraint on integer distortion
$\widehat{\mathrm{F}}_{\!M}(m) = \frac{1}{m + 1}$. We choose this word
size of $L=3$ in keeping with the hand-worked examples in the
previous sections. Because we have an automated tool
(Section~\ref{section:howWeSolveUsingDreal}) to generate the code
tables as well as to perform the Monte Carlo simulation, our methods
and validation are not limited to small word sizes.

We perform the Monte Carlo simulation with 10,000 trials, where
each trial picks a value to be transmitted uniformly from the set
of possible $L$-bit values and induces $k$ upsets with probability
determined by the encoding.  We then compare the mean resulting
integer distortion with the constraint $\widehat{\mathrm{F}}_{\!M}(m)$.

\begin{figure}
\centering
\subfloat[When channel errors in all ordinal positions of words are i.i.d, the code table is a single permissible channel error probability, $p$.]{\includegraphics[trim=0cm 0cm 0cm 0cm, clip=true, angle=0, width=0.235\textwidth]{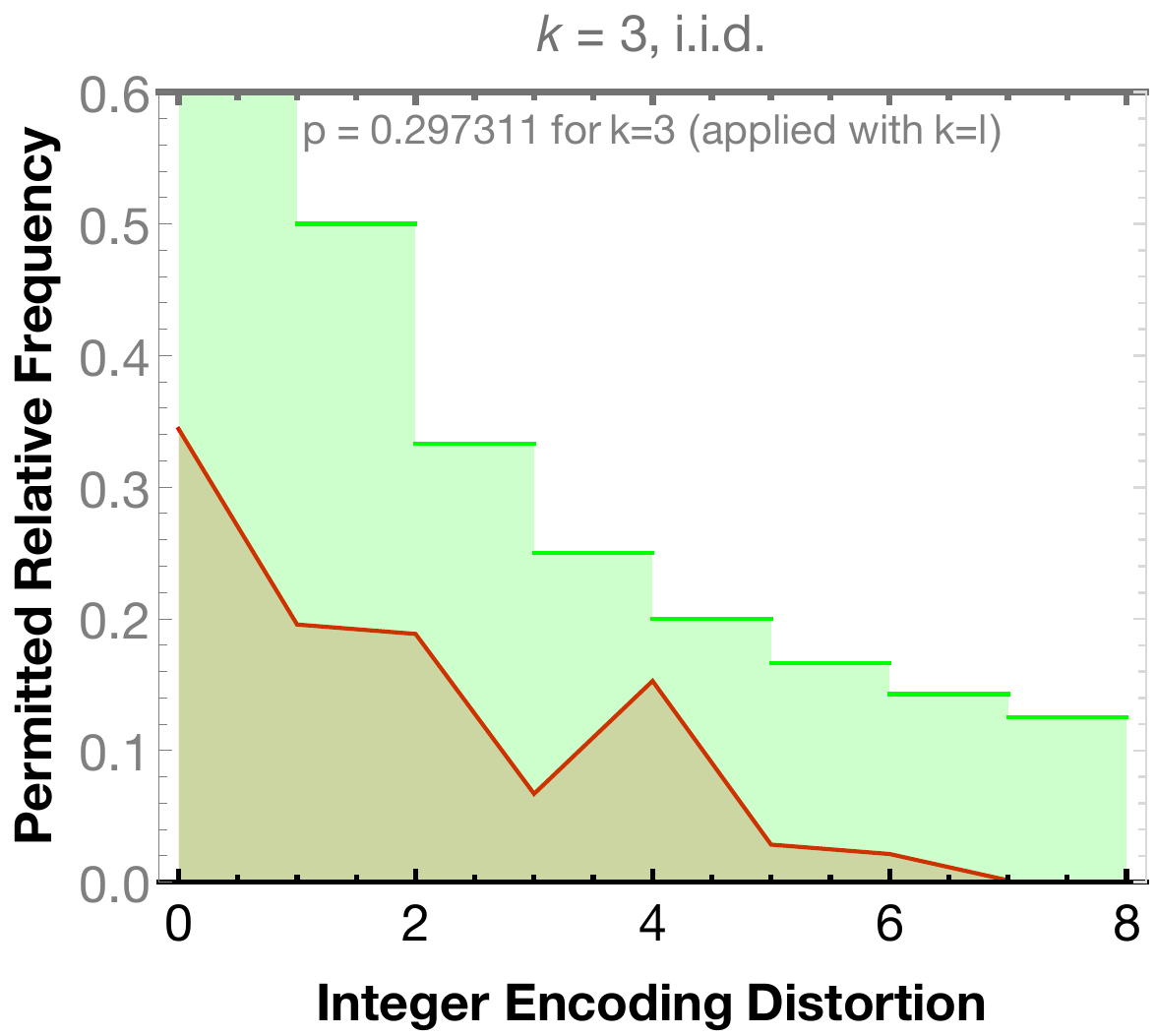}}~~
\subfloat[When channel errors are not identically distributed, the code table is a different channel error probability, $p_i$, for each of the $L$ bits.]{\includegraphics[trim=0cm 0cm 0cm 0cm, clip=true, angle=0, width=0.235\textwidth]{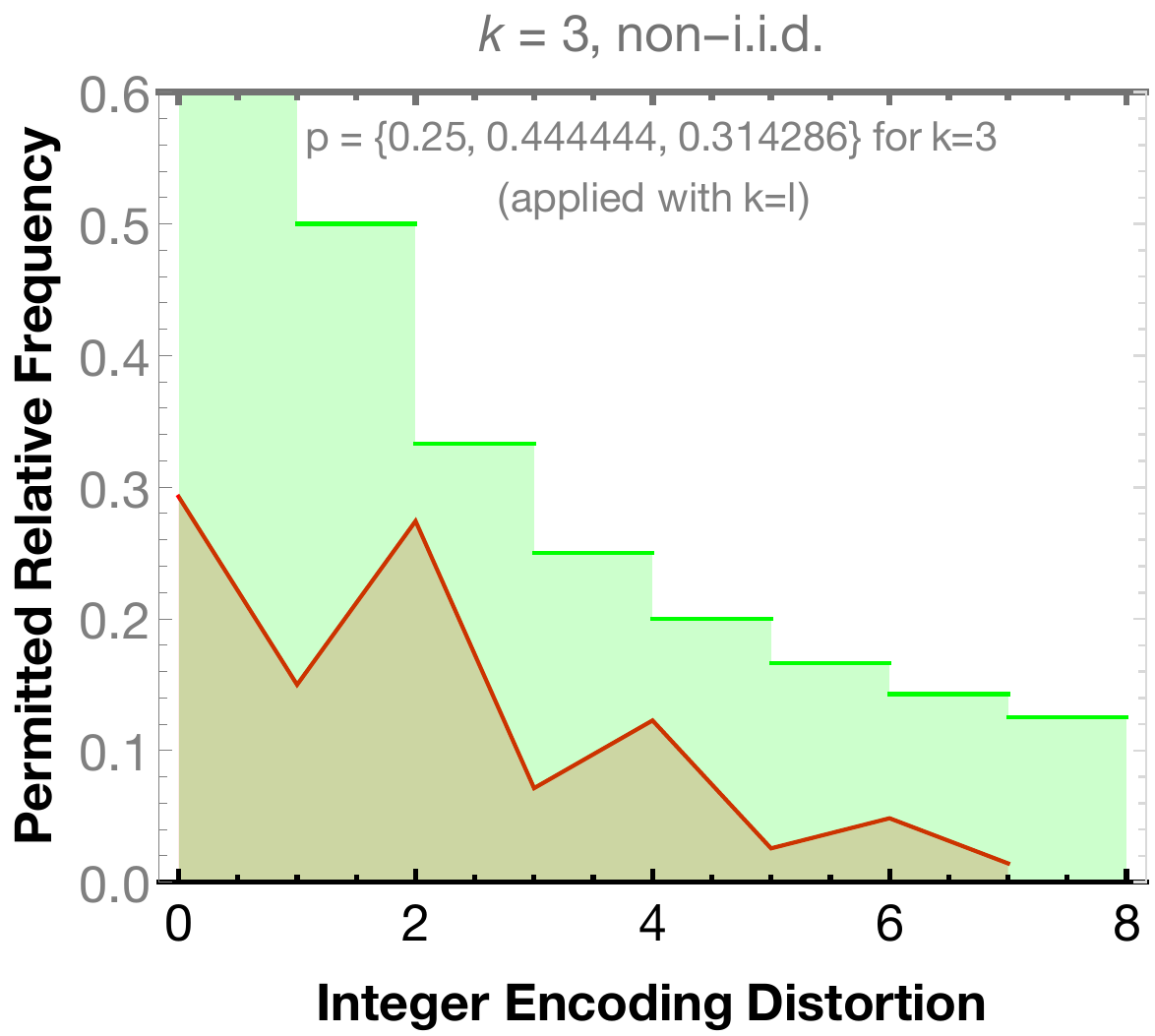}}
\caption{The distribution of actual induced integer distortion (dark/red)
is strictly within the bounds dictated by the constraint $\widehat{\mathrm{F}}_{\!M}(m)$ (light/green).}
\label{fig:Fm-and-FmHat-L3-k3}
\end{figure}

Figure~\ref{fig:Fm-and-FmHat-L3-k3}(a) shows the results for the
i.i.d. scenario where all bits of a word are forced to be in error
with the same probability.  The method described in
Section~\ref{section:iidBernoulli} determines the permissible
channel error probability to be $p = 0.29$.  As
Figure~\ref{fig:Fm-and-FmHat-L3-k3}(a) shows, the distribution of
actual induced integer distortion (dark/red) is strictly within the
bounds dictated by the constraint (light/green). 
Figure~\ref{fig:Fm-and-FmHat-L3-k3}(b) shows the results for the
non-i.i.d. scenario. In this case, the method of
Section~\ref{section:nonIidErrors} results in a code table with
the $p_i$s for the $L=3$ bits being $\{0.25, 0.44, 0.31\}$.

\section{Hardware implementation}
\label{section:hardware}
The energy required to transmit a bit is typically a monotonic
function of the channel bit-error rate (BER). Examples of methods
by which the energy per bit and hence the BER may be controlled
include varying the transmission bit rate, varying the I/O voltage
swing, and varying I/O termination\iftoggle{pnasTemplate}{ (Figure~\ref{fig:termination}).}{.}

\iftoggle{pnasTemplate}{
\begin{figure}
\centering
{\includegraphics[trim=0cm 0cm 0cm 0cm, clip=true, angle=0, width=0.425\textwidth]{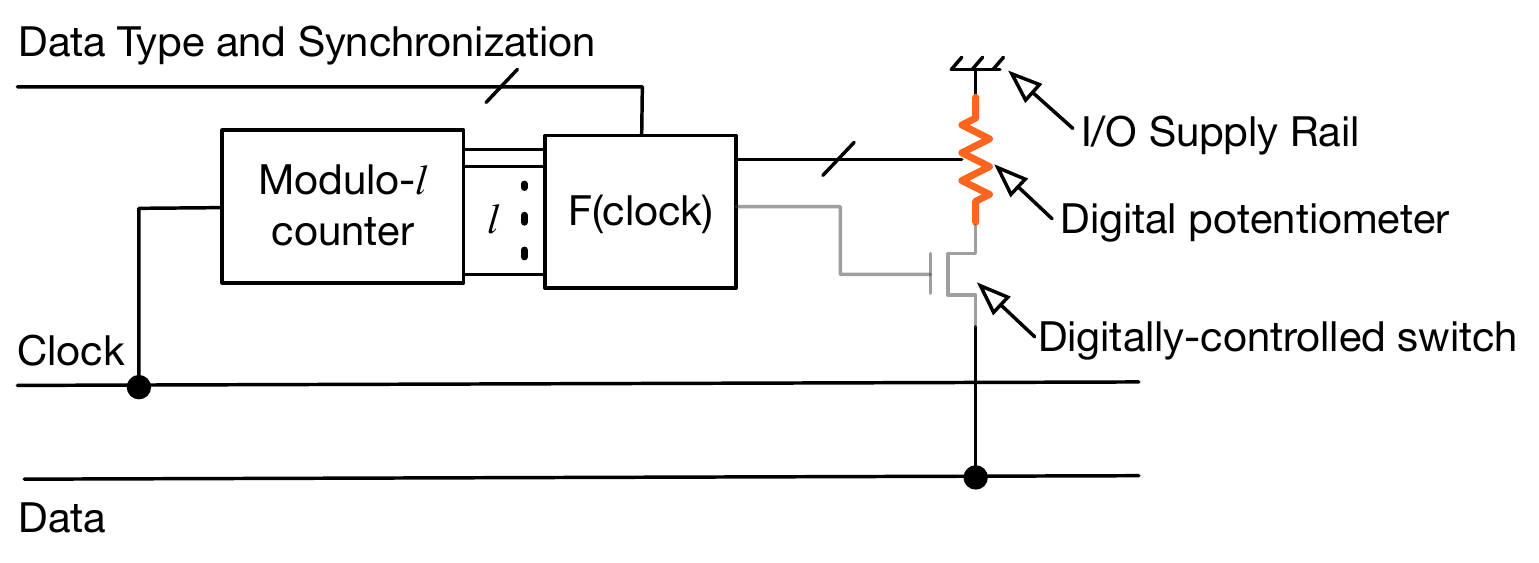}}
\caption{The function F(clock) outputs a control signal for the
digitally-controlled termination potentiometer, with the termination
value chosen so that at a given communication data rate, bit errors
occur with with probability $p_i$ for the $i$'th bit of a word,
corresponding to the bits which are transmitted with reduced
reliability. A degenerate case in which the potentiometer value is
always $0$ or $\infty$ corresponds to a per-bit choice of completely
removing the pull-up resistor.}
\label{fig:termination}
\end{figure}
}
{}

\iftoggle{pnasTemplate}
{
	\begin{SCfigure}[\sidecaptionrelwidth][t]
		\centering
		{\includegraphics[trim=0cm 0cm 0cm 0cm, clip=true, angle=0, width=0.3\textwidth]{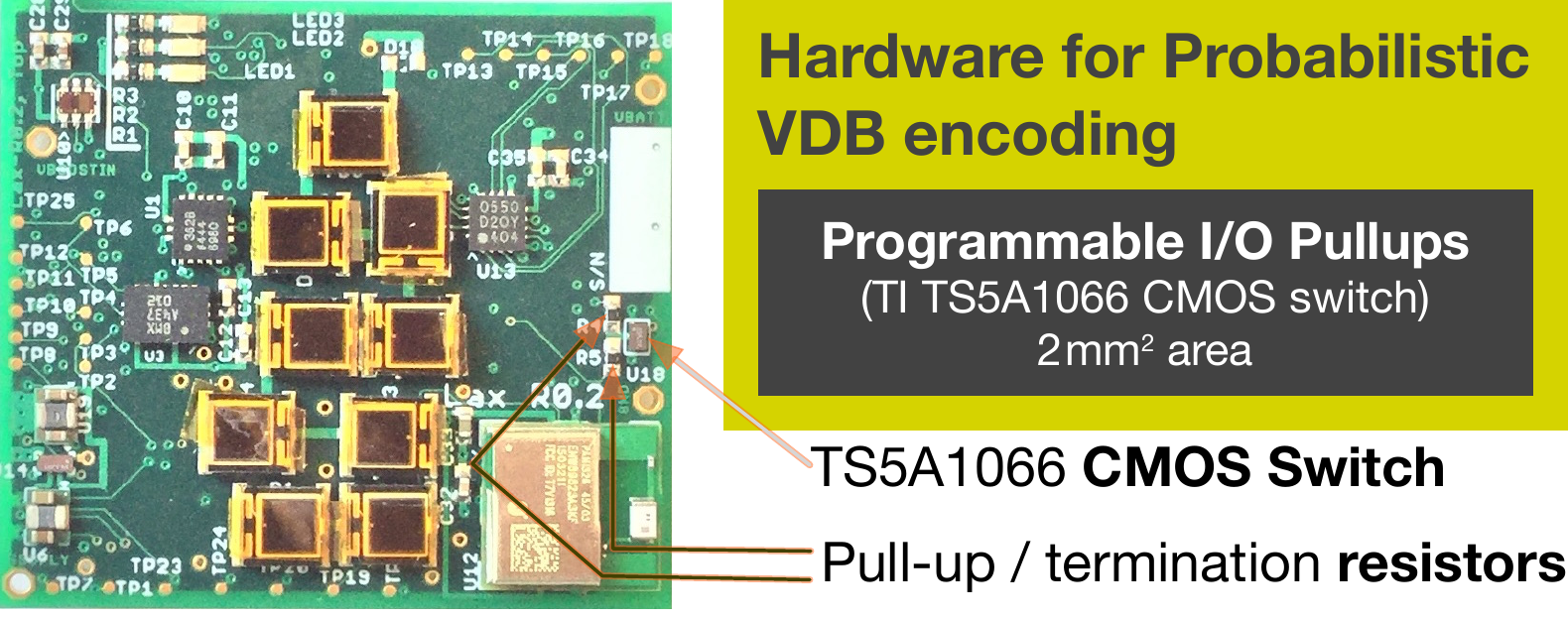}}
		\caption{Implementation of a variant of the technique from
		Figure~\ref{fig:termination} in the \Warp hardware platform.
		The only hardware overhead is the TS5A1066 CMOS switch integrated circuit.}
		\label{fig:termination-in-warp}
	\end{SCfigure}
}
{
	\begin{figure}[t]
		\centering
		{\includegraphics[trim=0cm 0cm 0cm 0cm, clip=true, angle=0, width=0.3\textwidth]{ILLUSTRATIONS/warp-hardware-ts5a1066-for-VDB-encoding.pdf}}
		\caption{Implementation of a variant of the technique from
		Section~\ref{section:hardware} in the \Warp hardware platform.
		The only hardware overhead is the TS5A1066 CMOS switch integrated circuit.}
		\label{fig:termination-in-warp}
	\end{figure}
}

We implemented a variant of the programmable link termination
method\iftoggle{pnasTemplate}{from Figure~\ref{fig:termination}}{}
for an I2C communication interface.  I2C is a popular communication
interface for connecting sensors such as accelerometers and gyros
in embedded systems. We implement the programmable termination
resistor using a CMOS switch and integrated our implementation
within \Warp, a research platform for investigating accuracy versus
efficiency tradeoffs in embedded sensing systems. Our circuit
implementation uses an integrated circuit that occupies less than
2\,mm$^2$ in area and dissipates less than 0.5\,$\mu$W, in addition
to the termination/pull-up resistors which are already required in
designs (Figure~\ref{fig:termination-in-warp}).

\iftoggle{pnasTemplate}
{
The values of tolerable error probabilities from
Section~\ref{section:iidBernoulli} and Section~\ref{section:nonIidErrors}
enable communication channel interfaces where the energy expended
per bit is modulated by the tolerable bit error probability across
the bits of a word. Such communication interfaces provide reduced
I/O power dissipation, while incurring integer distortions of
transmitted values no worse than $\widehat{\mathrm{F}}_{\!M}(m)$.
}
{}

\section{Conclusions}
\label{section:conclusions}
\iftoggle{pnasTemplate}{
Computing systems which interact with the physical world typically
process noisy data. As a result, the algorithms which execute in
these computing systems can often tolerate some degree of deviation
from correctness of their inputs (in terms of integer- or real-valued
distortions as opposed to Hamming distortions in arbitrary
representational bit positions), without significantly affecting
their outputs. When the outputs of computing systems are for
consumption by humans (e.g., images) or used in contexts where there
is no strict reference of correctness (e.g., outputs of a step
counting pedometer algorithm), the tolerance to errors can be taken
advantage of to improve time efficiency (performance) or
energy-efficiency, by purposefully inducing errors. This article
introduces a \textit{probabilistic value-deviation-bounded codes}
(probabilistic VDB codes), a technique for probabilistically encoding
communication data to take advantage of these observations to reduce
the energy cost of data communication in embedded systems that use
I/O interfaces like SPI and I2C.
}{}

We present \iftoggle{pnasTemplate}{tighter upper bounds on the
efficiency for VDB codes and present}{} a new energy-saving
probabilistic VDB encoder that trades the integer distance in induced
errors for lower energy usage.  The code we present takes the
peculiar approach of changing the channel bit error rate across the
ordinal bit positions in a word to reduce power dissipation.  We
present two methods for generating the probabilistic VDB code tables:
one method is based on a straightforward enumeration of the code
space, and the second method builds on ideas from finite field
arithmetic and number theory to efficiently generate the code tables.
We present one realization of hardware to implement the technique,
requiring 2\,mm$^2$ of circuit board area and dissipating less than
0.5\,$\mu$W.

\acknow{This research is supported by an Alan Turing Institute award
TU/B/000096 under EPSRC grant EP/N510129/1, and by Royal Society grant RG170136.}

\showacknow %

\pnasbreak

\bibliography{probabilistic-vdbe}

\end{document}